\documentclass[aps,prl,twocolumn,superscriptaddress,groupedaddress,notitlepage]{revtex4-1}  

\usepackage{amsmath, amsfonts, amssymb}
\usepackage{graphicx}
\usepackage{enumerate}
\usepackage{color}
\usepackage{hyperref}
\usepackage{natbib}

\newcommand{\be}{\begin{equation}}
\newcommand{\ee}{\end{equation}}

\newcommand{\figref}[1]{Fig.~\ref{#1}}

\def\be{\begin{equation}}
\def\ee{\end{equation}}
\def\beq{\begin{equation}}
\def\eeq{\end{equation}}

\newcommand{\gam}{\gamma}

\newcommand{\del}{\delta}
\newcommand{\Del}{\Delta}
\newcommand{\eps}{\epsilon}

\newcommand{\Lam}{\Lambda}

\newcommand{\de}{\partial}
\newcommand{\rmd}{\mathrm{d}}
\newcommand{\nab}{\nabla}

\renewcommand{\[}{\left[}
\renewcommand{\]}{\right]}
\renewcommand{\(}{\left(}
\renewcommand{\)}{\right)}

\newcommand{\Mpl}{M_{\textrm{Pl}}}
\newcommand{\x}{\vec{x}}

 \def\simleq{\; \raise0.3ex\hbox{$<$\kern-0.75em
      \raise-1.1ex\hbox{$\sim$}}\; }
   \def\simgeq{\; \raise0.3ex\hbox{$>$\kern-0.75em
      \raise-1.1ex\hbox{$\sim$}}\; }

\begin{document}

\title{Non-Gaussianity after BICEP2}
\author{Guido D'Amico}
\email{gda2@nyu.edu, $^{\dagger}$  kleban@nyu.edu}
\author{Matthew Kleban$^{\dagger}$}
\affiliation{Center for Cosmology and Particle Physics, New York University, New York, NY}
\date{\today}

\begin{abstract}
We analyze  primordial non-Gaussianity in single field inflationary models when the tensor/scalar ratio is large.  Our results show that detectable levels of  non-Gaussianity $f_{NL} \sim 50$ are still possible in the simplest  class of  models described by the effective theory of inflation.  However, the \emph{shape} is very tightly constrained, making  a sharp prediction that could  be confirmed or falsified by a future detection of non-Gaussianity.
\end{abstract}

\maketitle

Current cosmological data provides strong support for the simplest models of inflation.  A single scalar field with a quadratic potential and canonical kinetic term ${\cal L} = -(\partial \phi)^{2}/2 - m^{2} \phi^{2}/2$~\cite{Linde:1983gd, Kaloper:2008fb, Kaloper:2011jz, Creminelli:2014oaa} predicts a tilt $n_{s}-1$ and tensor/scalar ratio $r$ that lie within about one sigma of the current central values reported by Planck and BICEP2~\cite{Ade:2013zuv, Ade:2014xna}.
It also predicts  extremely Gaussian perturbations, with undetectably small non-Gaussian deviations  $f_{NL} \sim \epsilon$~\cite{Maldacena:2002vr}.  

Planck improved the constraints on non-Gaussianity in the CMB temperature spectrum~\cite{Ade:2013ydc} (\figref{plot}).  In addition the recent BICEP2 results indicate that the tensor/scalar ratio $r \sim 0.2$.  At the level of inflationary model building,   such a large value of $r$ at least naively  poses  difficulties for detectable non-Gaussianity.  This is because the simplest way to achieve large $f_{NL}$ is an inflationary model with a non-canonical kinetic term such that the speed of sound of perturbations during inflation  $c_{s} \ll 1$.  This produces non-Gaussianity peaked on equilateral triangles,  with $f_{NL} \sim 1/c_{s}^{2}$ \cite{Gruzinov:2004jx, Cheung:2007st}.  But small $c_{s}$ also enhances the amplitude of scalar perturbations so that $r \propto c_{s}$, making it  difficult to obtain both large $r$ and large $f_{NL}$ in this way.  

In this note we will argue that in these ``next-to-minimal'' models of inflation -- namely, single-field models described by the effective theory of inflation~\cite{Cheung:2007st} up to cubic order in perturbations -  it is still possible to achieve relatively large (and detectable) levels of primordial non-Gaussianity even with $r \sim 0.2$.  This is because there are \emph{two} independent parameters that determine $f_{NL}$:  the speed of sound $c_{s}$, and another parameter $c_{p}$ (that appears in the coefficient of $\dot \pi^{3}$).  

As mentioned above $c_{s}$ is fairly tightly constrained if $r \sim 0.2$.  But $c_{p}$ is not, and, while $c_{s} \ll 1$  generates $c_{p} \gg 1$ via quantum loop corrections, the converse is not true.  In other words, an effective theory with $c_{s} \simleq 1$ but $c_{p} \gg 1$ is technically natural.  Furthermore it can arise in a regime where higher derivative terms and modifications of Einstein gravity are small.  

The main result we wish to report in this note is that not only a detectable $f_{NL}$ is possible in this ``next-to-minimal'' scenario, but its \emph{shape} is very tightly constrained (\figref{plot}).  This means -- assuming that the tensor/scalar ratio $r$ is in fact large -- that a detection of non-Gaussianity with a shape that does \emph{not} lie in the narrow region plotted in \figref{plot} would have dramatic implications for the theory of inflation and fundamental physics in general.

\section{Effective Theory of Inflation}
\label{sec:EFTI}

The effective theory of inflation is based on the fundamental organizing principle of effective field theory:   one should write an action containing all terms compatible with the underlying symmetries, starting from the terms most relevant at low energies.
In the case of  inflation the  symmetries are those of de Sitter spacetime, with time-translation invariance weakly broken by slow-roll corrections.
The appropriate metric to describe background plus fluctuations can be written in ADM form:
\be
\rmd s^2 = - N^2 \rmd t^2 + h_{ij} (\rmd x^i + N^i \rmd t) (\rmd x^j + N^j \rmd t) \, ,
\ee
where the background is quasi de Sitter: $N = 1$, $N^i = 0$, $h_{ij} = a(t)^2 \del_{ij} = e^{2 H t} \del_{ij}$, with $H$ a slowly varying function of $t$.
The effective action for perturbations of single field inflation, in unitary  gauge and to cubic order in perturbations, is
\begin{equation}
\label{a1}
\begin{split}
S &= \int \sqrt{-g} \bigg[ \frac{\Mpl^2}{2} R - \Mpl^2 (3 H^2 + \dot{H}) + \Mpl^2 \dot{H} \frac{1}{N^2} \\
&+ \frac{M_{2}^4}{2!} \( 1 - \frac{1}{N^2} \)^2 + \frac{M_3^4}{3!} \( 1 - \frac{1}{N^2} \)^3 \\
&- \frac{\bar{M}_1^3}{2} \(1 - \frac{1}{N^2} \) \del K - \frac{\bar{M}_2^2}{2} \del K^2
- \frac{\bar{M}_3^2}{2} \del K^i_j \del K^j_i + \dots  \bigg].
\end{split}
\end{equation}
Here $\del K^i_j = K^i_j - H \del^i_j$, where $K^i_j$ is the extrinsic curvature $K^i_j = h^{ik} \( \de_t h_{k j} - \nab_k N_j - \nab_j N_k\)$, and the $M_i, \bar M_i$ are constants with dimensions of mass.
The $\dots$ refers to terms of higher order in derivatives or fluctuations.  

The first line in \eqref{a1} incorporates all single-field models with canonical kinetic terms.  As mentioned above, these models produce negligible non-Gaussianity.  The terms in the second line arise from non-canonical kinetic terms, and can produce large $f_{NL}$.  The main result of this paper is that the Planck and BICEP2 data constrain these terms in such a way that the shape of non-Gaussianity they produce is very sharply defined (\figref{plot}). The terms in the third line come from higher derivative terms that cannot be removed by integration by parts, as in ghost condensation~\cite{ArkaniHamed:2003uy}, or perhaps from Lorentz-breaking modifications of gravity. These terms can produce non-Gaussianity with various different shapes, and -- because they indicate exotic physics -- experimental confirmation that they are important for the dynamics of inflation would be of great importance.

We introduce the Goldstone boson of broken time translations $\pi$ by making the diffeomorphism:
\be
t \to \tilde{t} = t - \pi(t, \x) \, , \qquad x^i \to x^i \, .
\ee
As we are interested in non-negligible non-Gaussianities ($f_{NL} \gg 1$), we can neglect metric perturbations.
\begin{equation}
\label{a3}
\begin{split}
S_{\pi} &= \int a^3 \bigg[ \( - \Mpl^2 \dot{H} + 2 M_2^4 \) \dot{\pi}^2
+ \Mpl^2 \dot{H} \frac{(\de_i \pi)^2}{a^2} \\
&-  2 M_2^4 \dot{\pi} \frac{(\de_i \pi)^2}{a^2} + (2 M_2^4 - \frac{4}{3} M_3^4) \dot{\pi}^3
- \frac{\bar{M}_1^3 H}{2} \frac{(\de_i \pi)^2}{a^2} \\
&- \frac{\bar{M}_2^2 +\bar{M}_3^2}{2} \frac{(\de_i^2 \pi)^2}{a^4}
+ \frac{\bar{M}_1^3}{2} \frac{(\de_i^2 \pi) (\de_j \pi)^2}{a^4} + \dots  \bigg],
\end{split}
\end{equation}
where again $\dots$ indicate terms of higher order in derivatives or fluctuations.

In general, for a scalar field $\pi$ in de Sitter space with  Lagrangian
\be \label{a2}
S = \int a^3 N_c \[ \dot{\pi}^2 - c_s^2 \frac{(\de_i \pi)^2}{a^2} \]
\ee
the speed of propagation is $c_{s}$ and we get the power spectrum
\be
\Del_\pi = \frac{1}{8 \pi^2} \frac{H^2}{N_c c_s^3},
\ee
where the dimensionless power spectrum is defined by $\Del = k^3 P(k)/(2 \pi^2)$.

Let us pause to briefly discuss tensors.
The tensor perturbations are defined by setting
\be
h_{ij} = a(t)^2 (e^\gam)_{ij}
\ee
with $\det \exp(\gam) = 1$, $\de_i \gam_{ij} = 0$.
The kinetic term for $\gamma$ comes from the Einstein-Hilbert term in \eqref{a1}, with a correction from the $\del K^i_j \del K^j_i$ term: 
\be
S_{\gam} = \frac{\Mpl^2}{8} \int a^3 \[ \(1 - \frac{\bar{M}_3^2}{\Mpl^2} \) \dot{\gam}_{ij}^2
- \frac{(\de_k \gam_{ij})^2}{a^2} + \dots \].
\ee
This shows  there is a modification of the tensor speed of sound, $c_T^{-2} = 1 - \bar{M}_3^{2}/M_{Pl}^2 $.
This  differs significantly from $c_{T}=1$ only if $\bar{M}_3^2$ is close to the Planck mass.  Even in this case the resulting tensor non-Gaussianity will 
be only ${\cal O}(1)$ enhanced compared to its ordinary (undetectably small) value.  For this reason we  ignore tensors as a source for non-Gaussianity.   The tensor power spectrum is given by
\be
\Del_T = \frac{2}{\pi^2} \frac{H^2}{c_T \Mpl^2}.
\ee

\section{ ``Next-to-minimal'' models of inflation}

The class of models with $\bar{M}_i = 0$ arise from  theories with a non-canonical kinetic term $P(X)$ for a scalar field $\phi$, where $P$ is a function and $X \equiv (\partial \phi)^{2}$.
With $\bar{M}_i = 0$, eq.~\eqref{a3} simplifies to
\be \label{ntm}
\begin{split}
S_{\pi} &= \int a^3 \frac{\Mpl^2 |\dot H|}{c_s^2}
\bigg[ \( \dot{\pi}^2 - c_s^2 \frac{(\de_i \pi)^2}{a^2} \) \\
&- (1-c_s^2) \dot{\pi} \frac{(\de_i \pi)^2}{a^2}  
+ \(1-c_s^2 + \frac{2}{3} c_p \) \dot{\pi}^3 \bigg] \, ,
\end{split}
\ee
where we defined $c_s^2 \equiv \Mpl^2 \dot{H}/(\Mpl^2 \dot{H} - 2 M_2^4)$ and $c_p = 2 M_3^4/(\Mpl^2 \dot{H} - 2 M_2^4)$, so that $c_p$ stays well defined in the limit $M_2 \to 0$.  For  reference, the parameter $c_{3} \equiv - M_3^4/M_2^4$ is defined by $c_p = c_3 (1-c_s^2)$.
The tensor-to-scalar ratio in this model is
\be
r = 16 \eps c_s \, .
\ee
Since $n_{s}-1 \sim \epsilon$ and Planck measured $n_{s} \sim .96$~\cite{Ade:2013zuv},   one  concludes that $c_s$ cannot deviate substantially from $r$:
\be \label{cs}
c_s \simgeq r.
\ee
The non-Gaussianity induced by the operator $\dot{\pi} (\de \pi)^2$ is proportional to $(1-c_s^{-2})$, and therefore cannot be very large if $r$ is large.
However, the term ${2 \over 3} c_{p} \dot{\pi}^3$ can still give a sizable contribution to the non-Gaussianity if $c_p \gg 1$.  

Quantum loops in the effective theory generate $ \Delta M_{3}^{4} \sim M_{2}^{4}/c_{s}^{2}$~\cite{Senatore:2010jy, Baumann:2011nk}.
Hence, absent fine-tuning, $c_{s} \ll 1$ generates $c_p \sim c_{s}^{-2} \gg 1$.  One might expect the converse to hold as well: that $c_p \gg 1$ generates small $c_{s}$.
However this is not the case.
Up to cubic order there is no diagram that generates $\dot{\pi} (\de \pi)^2$ from $\dot \pi^{3}$.
Therefore,  $c_p \gg 1$ with $c_{s} \sim 1$ is technically natural~\footnote{Expanding \eqref{a3} to \emph{quartic} order in fluctuations gives a term $\sim c_p \dot \pi^{2} (\de \pi)^2$, which together with $\dot \pi^{3}$ generates $\dot{\pi} (\de \pi)^2$, but this is suppressed both by a loop factor and the fact that it arises from a quartic term.
Another way of seeing the same thing is to compute the 1-loop wave-function renormalization of $\pi$ due to the interaction $\dot \pi^{3}$. This gives rise at most to $\delta c_{s} \sim 1$.}~\cite{Creminelli:2014oaa}.

\begin{figure}[ht]
\includegraphics[scale=0.5]{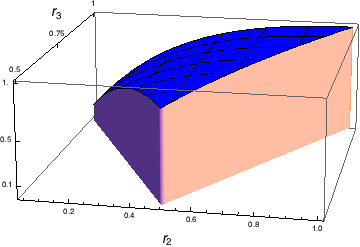}
\caption{ \label{shape} Shape of the bispectrum induced by $\dot{\pi}^{3}$.
As usual, we show $B(1,r_2^2,r_3^2) r_2^2 r_3^2$ where $r_i = k_i/k_1$ and we normalize to 1 in the equilateral limit.}
\end{figure}

In \figref{shape} we plot the shape of the bispectrum induced by the $\dot{\pi}^3$ operator, as a function of the two independent ratios of the triangle sides. Notice that this shape has a large flattened limit (triangles with $k_2 = k_3 = k_1/2$).

\begin{figure}[ht]
\includegraphics[scale=0.6]{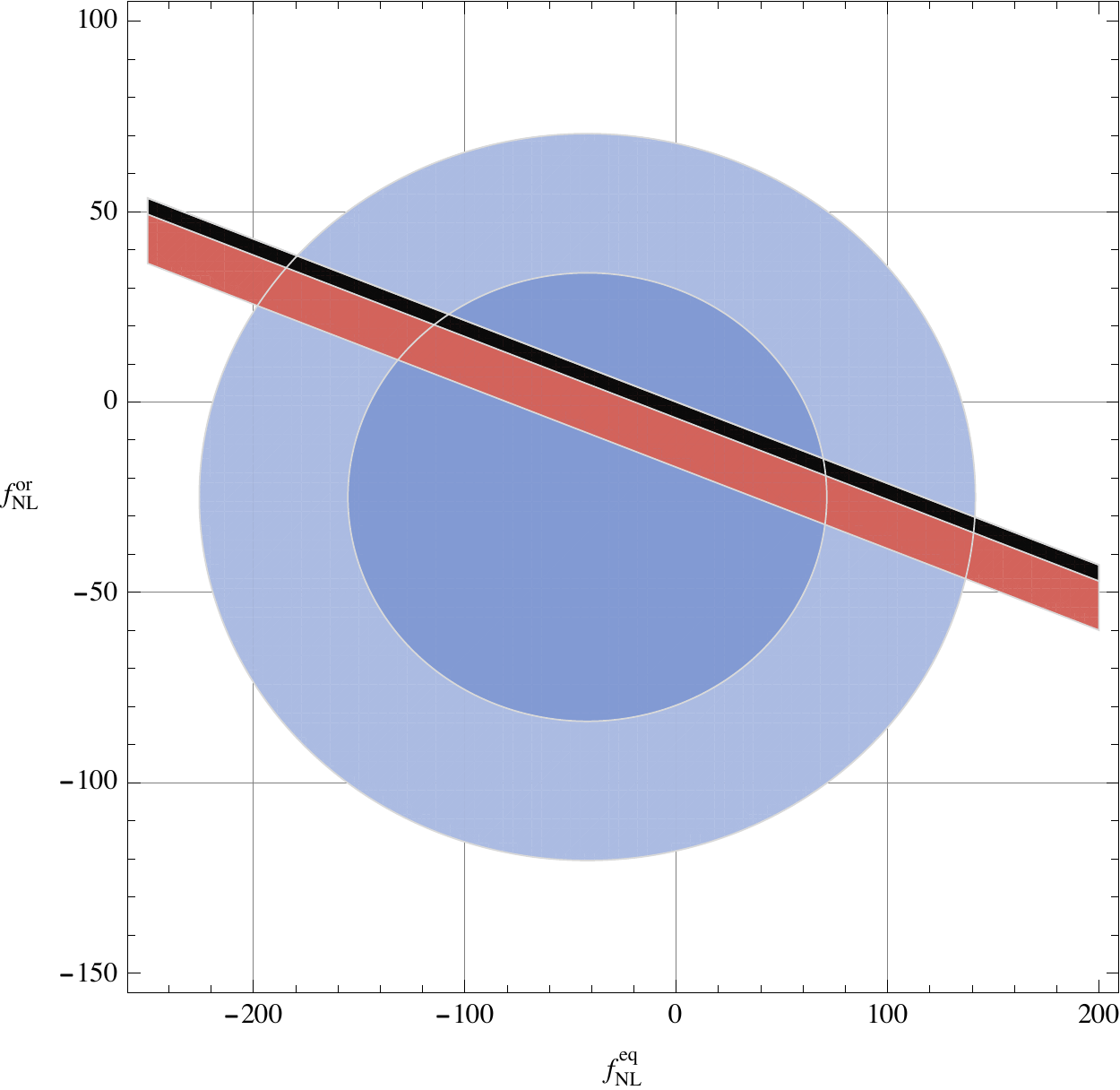}
\caption{ \label{plot} Constraints on $f_{NL}$ in the ``next-to-minimal'' model of inflation \eqref{ntm} for large $r$ \eqref{cs} for the equilateral/orthogonal plane, assuming  $c_s > 0.1$ (narrow black strip) or $c_{s} > .05$ (wider red strip), with the 1- and 2-$\sigma$ Planck constraints superimposed (circular blue regions).}
\end{figure}

In \figref{plot} we plot the possible values of $f_{NL}$ in the equilateral/orthogonal plane~\cite{Senatore:2009gt} in \eqref{ntm}, with the Planck constraints superimposed~\cite{Ade:2013zuv}.
As is evident, $r \sim 0.2$ (and therefore a large speed of sound) puts a very sharp constraint on the possible shape of non-Gaussianity in this class of models.
In the  ``next-to-minimal'' models the region above the line in \figref{plot} can be reached only allowing \emph{negative} $c_s^2$, and as such it would signal exotic physics even if $r$ is small.

\paragraph*{Dirac-Born-Infeld inflation:}
The case of DBI inflation~\cite{Alishahiha:2004eh} is $c_p = -3 (c_s^{2} -1)^{2}/2 c_{s}^{2} $.
The non-Gaussianity is $f_{NL}^{DBI} = \frac{105}{324} (1 - c_s^{-2}) \simgeq -30$ if $c_s \simeq 0.1$, from the measurement of tensor modes.

\section{Other possibilities}
\label{sec:loopholes}

If a future experiment detects primordial non-Gaussianity  with a shape that does not fall in the narrow region indicated in \figref{plot}, what are the implications?  In this section we enumerate the set of models consistent with such an observation.  All of them differ sharply from the simple model considered above, and correspond to novel physics.

\subsection{Quasi-de Sitter}

If the background evolution is such that $\dot{H} \to 0$, the standard gradient kinetic term for the perturbations vanishes.
In this case, the additional terms in the effective action become important.  
To simplify the analysis, we  consider two limiting cases, depending on whether $d_{1}$ or $d_{2}+d_{3}$ dominates, where $d_1 \equiv 2 \bar{M}_1^3/M^3$, and $d_{2,3} \equiv \bar M_{2,3}^{2}/M^{2}$.  All the models in this class could potentially be tested by their prediction of near-exact scale invariance for tensor perturbations.

\paragraph*{$d_1 \gg d_{2} + d_{3}$:}

This case corresponds to a standard two-derivative kinetic term.
The $\pi$ Lagrangian is
\begin{equation}
\begin{split}
S_{\pi} &= \int a^3 \bigg[ 2 M^4 \(\dot{\pi}^2 - \frac{d_1 H}{8 M} \frac{(\de_i \pi)^2}{a^2} \)
-2 M^4 \dot{\pi} \frac{(\de_i \pi)^2}{a^2} \\
&+ 2 M^4 \(1 - \frac{2}{3} c_3 \) \dot{\pi}^3
+ \frac{d_1 M^3}{4} \frac{(\de_i^2 \pi) (\de_j \pi)^2}{a^4}  \bigg].
\end{split}
\end{equation}
The last term seems to imply an additional shape of non-Gaussianity.
However,  using the linear equations of motion and integrating by parts, it can be rewritten in terms of the standard one-derivative cubic terms:
\begin{equation}
\begin{split}
S_{\pi} &= \int a^3 2 M^4 \bigg[ \dot{\pi}^2 - c_s^2 \frac{(\de_i \pi)^2}{a^2} \\
&+ \dot{\pi} \frac{(\de_i \pi)^2}{a^2} + \(1 + \frac{2}{3} c_3 + \frac{2}{c_s^2} \) \dot{\pi}^3  \bigg] \, ,
\end{split}
\end{equation}
where the sound speed now is $c_s^2 = \frac{d_1 H}{8 M}$.
The interesting fact is that the tensor-to-scalar ratio $r$ now does not put strong constraints on the speed of sound:
\be
r = 32 \frac{M^4}{\Mpl^2 H^2} c_s^3 = \frac{8 M^2}{\pi \Mpl^2} \frac{c_s^{3/2}}{A_s}.
\ee
The requirement that the cutoff has to be $\Lam \gg H$ gives the constraint $c_s \gg A_s^{2/11} \simeq 0.03$.
The resulting non-Gaussianity falls in the upper right corner of the $f_{NL}^{eq} - f_{NL}^{or}$ plane, above the line in \figref{plot}.

\paragraph*{$d_1 \ll d_{2} + d_{3}$:}

This case corresponds to the ghost inflation regime~\cite{ArkaniHamed:2003uz}. The $\pi$ Lagrangian has a non-relativistic scaling for the kinetic term, which results in non-standard scaling dimensions for energy, momentum, and fields.
The most relevant operators are:
\begin{equation}
\begin{split}
S_{\pi} &= \int a^3 \bigg[ 2 M^4 \dot{\pi}^2 - \frac{(d_2 + d_3) M^2}{2} \frac{(\de_i^{2} \pi)^2}{a^4} \\
&-2 M^4 \dot{\pi} \frac{(\de_i \pi)^2}{a^2} + \frac{d_1 M^3}{4} \frac{(\de_i^2 \pi) (\de_j \pi)^2}{a^4}  \bigg] \, .
\end{split}
\end{equation}
  The cubic operators proportional to $d_2$, $d_3$ are suppressed by $H/M$.
In this situation, the tensor-to-scalar ratio is
\begin{equation}
\begin{split}
r &= \frac{8 M^2}{\pi \Mpl^2 c_T} \Gamma^2(1/4) \( \frac{M}{H} \)^{1/2} \(\frac{d_2+d_3}{4}\)^{3/4} \\
& \simeq 282 \frac{M^2}{\Mpl^2} \frac{(d_2+d_3)^{3/5}}{c_T} \, .
\end{split}
\end{equation}
Again, $r \sim 0.2$ does not impose strong constraints on the non-Gaussianity, which is very close to the equilateral template.

Notice that in the original work on ghost inflation~\cite{ArkaniHamed:2003uz}, tensor modes are taken to be negligible  because of constraints on the scale $\bar{M_2}+ \bar{M_3}$ from solar system tests of gravity.
However, in the effective theory approach, it is perfectly possible that the Lagrangian describing the inflationary perturbations is different from the one describing modifications of gravity today, so this constraint may not play a role.
For a recent discussion, see~\cite{Ivanov:2014yla}.

\subsection{Negative speed of sound}

It is possible to have consistent models in which $\dot{H} > 0$, which violate the null energy conditions and give a blue tensor tilt.
The speed of sound squared will be negative, which naively implies an instability of the theory.
However, the higher derivative gradient kinetic term $(\de_i^2 \pi)^2$ will make the theory stable at short length scales, which renders these models viable.
Depending on the relative size of the two gradient terms, the non-Gaussianity can be very small or extremely large (and already ruled out), but current constraints on $r$ do not put a stronger bound.

\subsection{Higher-derivative operators}

It is possible to impose a ``Galilean'' symmetry on the effective theory of perturbations which suppresses the single-derivative cubic interactions~\cite{Creminelli:2010qf}.
In this case, one has a standard kinetic term (with unit speed of sound), and a standard tensor-to-scalar ratio $r = 16 \eps$.
However, there can be sizable non-Gaussianity induced by 5-derivative cubic terms.
The shape of the resulting 3-point function is peculiar, being enhanced in the flattened configurations and somewhat suppressed in the equilateral one (see ~\cite{Creminelli:2010qf} for details).
Also, because of the different scaling of the relevant operators, the size of the 4-point function scales as $NG_4 \simeq NG_3^{8/5}$ instead of the usual $NG_3^2$.

\subsection{Modified tensor spectrum}

As mentioned above, if $\bar M_3^2/M_{Pl}^2 \approx 1$ it is possible to have $c_T \ll 1$.  Generically this makes the term $(\bar M_{2}^{2}+\bar M_{3}^{2})(\partial_{i}^{2} \pi)^{2}$ in \eqref{a3} very important, which strongly affects the scalar spectrum.  However if $\bar M_{2}^{2}+\bar M_{3}^{2}$ is tuned to be small or zero, the effect on scalars is small.  In this case since $r \sim c_s \epsilon /c_T$, if $c_{T} \ll 1$ the tensor/scalar ratio $r$ can be large even with $c_s \ll 1$~\cite{Noumi:2014zqa},  removing the constraint that allowed us to make a sharp prediction for the shape of non-Gaussianity.

\subsection{Beyond single-field}

Models with multiple fields are much less well constrained.
For instance, a model  in which a small fraction of the perturbations are generated by a curvaton can  evade isocurvature constraints, but if the fluctuations of the curvaton are highly non-Gaussian this could produce relatively large $f_{NL}$.  Another class of examples are dissipative models such as~\cite{Green:2009ds, LopezNacir:2011kk, DAmico:2013iaa}, which produces flattened non-Gaussianity.  Yet another possibility is quasi-single field inflation~\cite{Chen:2009zp}, whose characteristic is a non-trivial squeezed limit, intermediate between the local and equilateral forms.

\section{Conclusions}

While the very simplest models of inflation predict undetectably small non-Gaussianity, our analysis shows that the ``next-to-minimal'' models can produce detectable $f_{NL}$, but with a shape that is very tightly constrained by BICEP2's measurement of $r$.  Therefore, a measurement of non-Gaussianity with a shape that is inconsistent with this constraint indicates the presence of important new physics:  either multiple light fields that strongly affect the inflationary dynamics, or some novel type of higher derivative coupling or violation of Lorentz invariance.  

\begin{acknowledgments}
It is a pleasure to thank Paolo Creminelli, Andrei Gruzinov, Nemanja Kaloper, Diana L—pez Nacir, Enrico Pajer, Marko Simonovi\'c, Grabriele Trevisan, Filippo Vernizzi, and Matias Zaldarriaga for useful discussions.
The work of G. D'A. is supported by NASA through grant NNX10A171G and by NSF through AST-1109432.
The work of MK is supported in part by the NSF  through grant PHY-1214302 and by the John Templeton Foundation.  The opinions expressed in this publication are those of the authors and do not necessarily reflect the views of the John Templeton Foundation.
\end{acknowledgments}

\bibliography{bibliography.bib}

\end{document}